\documentclass[preprint,12pt]{elsarticle}
\biboptions{sort&compress}

\usepackage{epsfig}

\usepackage{amssymb}
\usepackage{graphicx}
\usepackage{subcaption}
\usepackage{xcolor}

\journal{Physics A}

\begin{document}
\newcommand{\etal}{{\em et al}. }
\newcommand{\ie}{{\em i. e.} }
\begin{frontmatter}



\title{Correlations between two vortices in dry active matter}


\author[label1]{Felipe P. S. Júnior}
\affiliation[label1]{organization={Universidade Federal do Pará},
            addressline={Faculdade de Física, ICEN, Av. Augusto Correa, 1, Guamá},
            city={Belém},
            postcode={66075-110},
            state={Pará},
            country={Brazil}}

 \author[label2]{Jorge L. C. Domingos}
\affiliation[label2]{organization={Universidade Federal do Ceará},
            addressline={Departamento de Física,  Caixa Postal 6030, Campus do Pici}, 
            city={Fortaleza},
            postcode={60455-760}, 
            state={Ceará},
            country={Brazil}}
            
\author[label1]{F. Q. Potiguar}

 \author[label2]{W. P. Ferreira}
\begin{abstract}
It was recently shown that wet active matter can form synchronized rotating vortices in a square lattice, similar to an antiferromagnetic Ising model (by considering rotation direction as spin projection). In this study, we investigate whether such a correlated state occurs for a model of dry active matter. We achieve that by numerically simulating the dynamics of a system of active particles in the presence of two identical circular obstacles. Then, we  measure the angular velocity correlation function of both vortices as a function of the obstacle diameter, their shortest separation (gap), and the particle density. 
When the correlation function is negative,
both vortices rotate in contrary directions. They maintain this state by exchanging particles through the region between them, analogously to synchronized cogs. On the other hand, with a positive correlation function, a single rotating cluster emerges, and the particles move around the whole structure, similar to a belt strapped around the obstacles. Additionally, we observe the emergence of uncorrelated states at the transition between correlated states, in which only a single vortex is present, or in the large gap regime, in which the vortices are nearly independent on each other.
\end{abstract}




\end{frontmatter}


\section{Introduction}\label{sec1}
 In active matter, particles generate their motion either by consuming their internal energy or by extracting the energy from the environment \cite{marchetti2013hydrodynamics,ribeiro2020trapping,das2020introduction,cates2010arrested}. 
 A collection of active particles is typically described using two distinct models: dry models, in which particles do not move under the influence of hydrodynamic interactions; and wet models, where hydrodynamic interactions are taken into account. We can also consider dry models as those where momentum is not conserved at the particle level. In these models, particles move in a fluid that does not affect their motions, apart from viscous friction,  resulting in overdamped behavior. In contrast, wet models involve strong fluid-particle the fluid interactions and thus, the total momentum (of both particles and fluid) is conserved \cite{marchetti2013hydrodynamics}.
Dry models can still be divided according to the particle-particle interaction, which can either include particle alignment interactions, {\em e.g.}, Vicsek model \cite{vicsek1995novel}, or not include such interactions, as Angular Brownian Motion (ABM) \cite{fily2012athermal}, which has a close relationship to Run-and-Tumble dynamics \cite{cates2013active}.

Active systems exhibit complex non-equilibrium phenomena and, although much has been achieved in understanding them, there still situations which need more investigations. Among several interesting properties of active systems, we can highlight the tendency to aggregate either spontaneously, where in free space the system exhibits spontaneous phase separation \cite{fily2012athermal}, or around rigid surfaces \cite{pan2020vortex,PhysRevE.101.022601,fily2014dynamics,spagnolie2015geometric,mokhtari2017collective,takagi2014hydrodynamic,qian2021porus,volpe2011microswimmers}; in the latter, it was shown that the aggregation process depends strongly on the local curvature \cite{fily2014dynamics}. Correlated to this property, there is the possibility of an active aggregate around a circular obstacle to rotate, forming the so-called active vortex, even in the absence of an external drive. Such a possibility was first shown in \cite{potiguar2014self}. Later, Mokhtari {\em et al.} \cite{mokhtari2017collective} described a particle capture-and-release mechanism through which a vortex persists for extended periods; Pan {\em et al}. \cite{pan2020vortex} performed a deeper investigation of the dynamics of a single vortex and observed that the obstacle diameter is a key parameter that controls the vortex dynamical regime. Finally, B. Qian {\em et al}. \cite{qian2021porus} reported an apparent rotation of a large aggregation of self propelled particles around lattices of tiny obstacles. Recently, it was reported the spontaneous emergence of vortices in the absence of obstacles which is intimately connected to the phenomenon of motility-induced phase separation (MIPS) typically seen in the ABP model \cite{caprini2020}; additionally, spontaneous rotation of clusters of dipolar \cite{liao2021} and attractive chiral particles \cite{caprini2024} were also reported.

On the other hand, for wet models, there is a richer literature addressing such a vortex formation, either experimentally \cite{takagi2014hydrodynamic,wioland2016ferromagnetic,beppu2017geometry} or numerically \cite{reinken2018derivation,beppu2017geometry}. The vortices are a consequence of mesoscale turbulence, in which the underlying fluid, perturbed by the active particles, drive them in a vortex-like pattern. In particular, in a regular square lattice of circular obstacles, Reinken {\em et al}.  \cite{reinken2022ising} observed correlated motions in neighboring vortices formed in the center of the unit cells of the original obstacle lattice, in which a pair of vortices rotate in contrary directions. Associating the mean normalized negative (positive) vorticity at each unit cell to a down (up) $1/2$-spin, they showed that this state displays an antiferromagnetic second-order transition with the strength of the nonlinear advection in the role of temperature, and critical exponents in the 2D Ising universality class. Finally, investigations of vortex interactions are also performed in wet models, for polar, active filaments \cite{gow16} and regular particles in free space \cite{rana20,rana22}, through solutions of the Toner-Tu equation \cite{toner95}, and a variety of dynamical phases, ordered and disordered, along with coarsening of the vortices is observed. 
Studies of vortex interactions in dry active models are still lacking.


In this work, we consider an ABM model to investigate correlations between two vortices as a function of the obstacles' diameter, particle density, and the shortest distance between the obstacles (gap). In particular, we measure the angular velocity correlation function of the vortices to characterize the correlated state. A correlated state refers to the synchronized rotational motion of vortices, where their angular velocities exhibit a defined relationship, either positive or negative.

We do not consider any external drive. We observe positive correlation states (PCS) and negative correlation states (NCS), which correspond, respectively, to a single large vortex around both obstacles and two vortices rotating in contrary directions. In the language of turbulence, they are counter-rotating vortices. We will show that the motion pattern for a NCS is similar to the one reported in \cite{reinken2022ising}, with the distinction that, here, the vortices appear around the obstacles and not in the center of a square cell of circular obstacles, in a way that allows us to call this state an antiferromagnetic state. However, the PCS, as we will show, does not correspond to two clearly isolated, co-rotating vortices, so that we do not call it a ferromagnetic state. Therefore, we refrain from using this terminology altogether.
We also characterize correlated, and consequently uncorrelated, states through the vertical particle current horizontal profiles, {\em i.e.}, as functions of the horizontal coordinate $x$, and, from an appropriate condition on this quantity, we define uncorrelated states and show that they appear at the transition between regimes or for large gaps.



\section{Model}
\label{model}
\begin{figure}[ht]	
 \centering
	\includegraphics[width=0.6\linewidth]{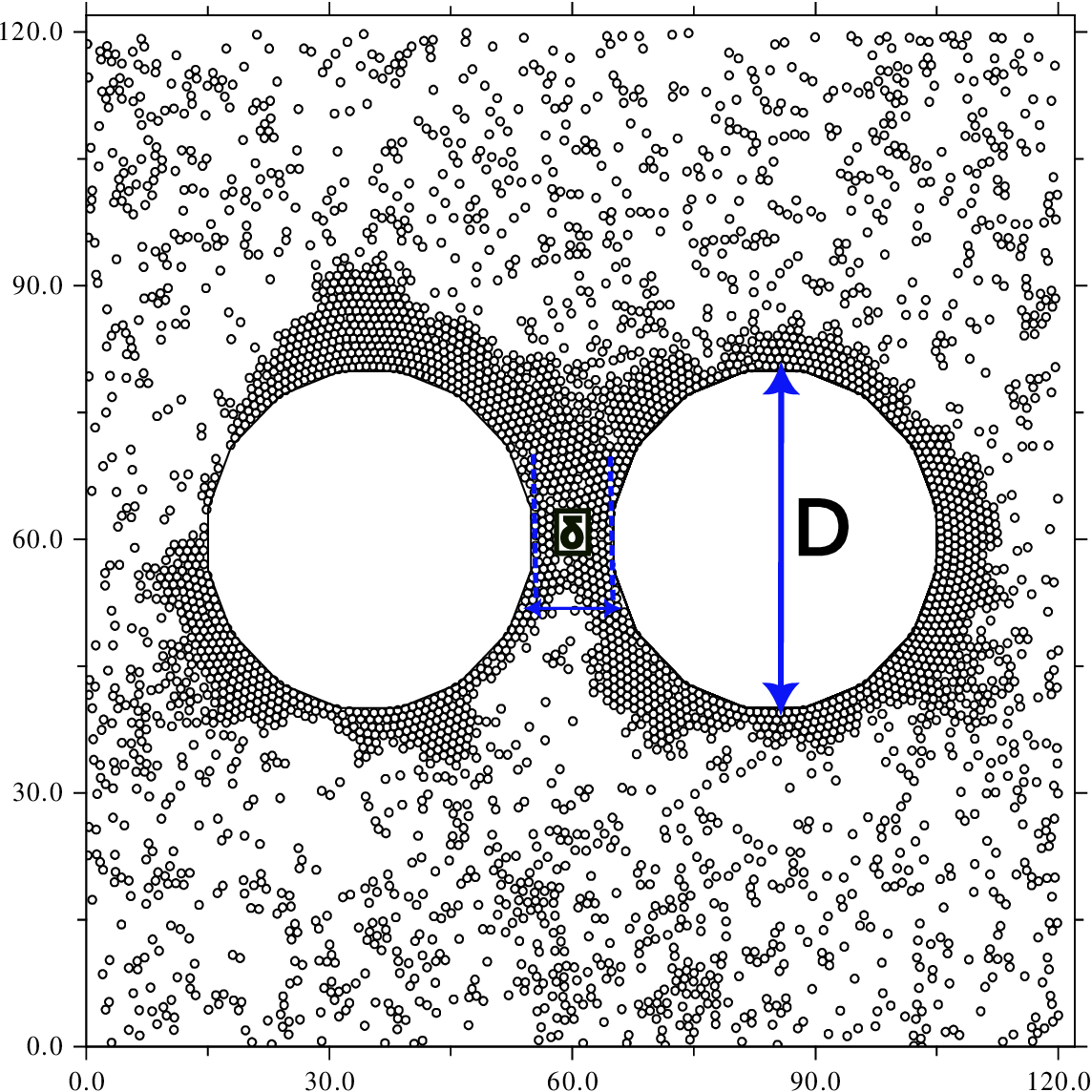}
	\caption{Typical particle configuration for our two obstacle system of diameter $D$ and separated by a gap $\delta$.}
	
	\label{fig:Top10}
\end{figure}

We consider a two dimensional ABM  \cite{fily2012athermal,marchetti2016minimal} model, composed of $N$ soft disks of diameter $\sigma$, and  an intrinsic velocity $v_0$ (self-propulsion).
The two identical circular obstacles, each with diameter $D$, are positioned such that their centers are located at $L/2\pm (\delta+D)/2$ along the x-axis. Here, $\delta$ represents the shortest distance between the obstacles, referred to as the gap (see Fig. \ref{fig:Top10}).
 
The disks interact through a linear spring
force law $\textbf{F}_{ij}=\kappa(d_{ij}-r_{ij})\hat{\textbf{r}}_{ij}$
for $d_{ij} > r_{ij}$, $\textbf{F}_{ij}=0$ otherwise, where $r_{ij}=|\textbf{r}_{i}-\textbf{r}_{j}|$ is distance between particles $i$ and $j$,  $\kappa$ is the spring stiffness, and 
$d_{ij}=\sigma$; for the disk-obstacle interaction, we use the same force law, but with $d_{ij}=\frac{1}{2}(\sigma+D)$ and  stiffness  $\kappa_{obs}=20\kappa$ in order to approach the rigid body limit.

The dynamics of the \textit{i}-th particle are given by the coupled Langevin equations
\begin{equation}
  \textbf{v}_{i}=\frac{d\textbf{r}_{i}}{dt}=\mu\textbf{F}_{i}+\textbf{v}^{f}_{i}+\textbf{A}_{i}(t) ,
\end{equation}
and
\begin{equation}
  \frac{d\theta_{i}}{dt}=\eta_{i}(t),
\end{equation}

\noindent
where  $\textbf{v}_{i}$ 
is the $i$-th particle velocity, $\mu$ is the motility, $\textbf{F}_{i}=\sum\limits_{j}\textbf{F}_{ij}$, where the sum runs over all particles and obstacles, is the total force in the \textit{i}-th particle, $  \textbf{v}^{f}_{i}=v_{0}(\cos\theta_{i}(t)\hat{\textbf{x}}+\sin\theta_{i}(t)\hat{\textbf{y}})$, is the intrinsic velocity with magnitude $v_{0}$, $\theta_{i}(t)$ is the angle with respect to the $x$-axis that sets the direction of the intrinsic velocity of particle $i$, while $\eta_{i}(t)$ is a delta-correlated white noise with zero average and unit variance, that follows:

\begin{equation}
\langle\eta_{j}(t_{2})\eta_{i}(t_{1})\rangle=2\eta_{\theta}\delta(t_{2}-t_{1})\delta_{ij},
\end{equation}

\begin{equation}
\langle\eta_{i}(t)\rangle=0.
\end{equation}
$\textbf{A}_{i}(t)$ is also a delta-correlated white noise variable with intensity $\xi$ that follows
\begin{equation}
\langle A_{\alpha j}(t_{2})A_{\beta i}(t_{1})\rangle=2\xi\delta(t_{2}-t_{1})\delta_{ij}\delta_{\alpha\beta},
\end{equation}

\begin{equation}
\langle\textbf{A}_{i}(t)\rangle=0,
\end{equation}
where, $\alpha$ and $\beta=x, y$. Given that our model is athermal, we set $\xi=0$. The brackets denote averages over independent realizations.

We used periodic boundary conditions (PBC) in both 
directions. The values of the model parameters are
$\sigma=1.0$ as the length unit, $\mu=1.0$, $L=120$, $\kappa=50.0$, $v_{0}=1$ (which also sets the time unit), $\eta_{\theta}=0.001$ and $\phi=\frac{N\pi\sigma^{2}}{4(L^{2}-\pi D^{2}/2)}$ as the particle density.

We calculate the angular velocity of vortex $\Gamma_{k}$ at time $t$ around obstacle $k$, centered on position $\textbf{R}_{k}$, $k=l,r$ for left and right obstacles, respectively, as
\begin{equation}
  \omega_{k}(t)=\sum_{i \in \Gamma_{k}}\frac{1}{N_{k}(t)}\frac{\textbf{v}_{i}(t)\cdot\hat{\psi}_{ik}(t)}{R_{ik}(t)} \label{Myomega},
\end{equation}
where $i\in \Gamma_{k}$ indicates that the sum runs over the particles $i$ that compose the vortex $\Gamma_{k}$ around the obstacle $k$, $R_{ik}(t)=|\textbf{r}_{i}(t)-\textbf{R}_{k}|$, $\hat{{\psi}}_{ik}(t)$ is the unit vector along the polar direction, as measured with the origin at the center of obstacle $k$, and $N_{k}(t)$ the number of particles in vortex $\Gamma_{k}$. We calculate this number with a cluster counting algorithm \cite{rapaport2004art} that identifies all clusters that touch obstacle $k$; if a particle belongs to such a cluster, it belongs to vortex $k$. If a cluster $k$ touches both obstacles, we split it into parts, one for each obstacle, and a particle belongs to the cluster around the obstacle to which the particle is closer. Using the angular velocities of both vortices 
we can define a global parameter, namely, the normalized angular velocity correlation function
\begin{equation}
    C_{\omega_{l}\omega_{r}}=
    \frac{\Big\langle\omega_{l}(t)\omega_{r}(t)\Big\rangle_t}{\Big\langle|\omega_{l}(t)\omega_{r}(t)|\Big\rangle_t}, \label{Cw}
\end{equation}
which is a measure of the correlation between the rotational motions of each vortex. The notation $<\cdot>_t$ indicates an average over time and independent realizations.
  
Our parameter space is $\delta=[2,..,20]$ with $\Delta\delta=2.0$ and $D=[25,...,40]$ with $\Delta D=5.0$, and area fractions $\phi=[0.2;0.25;0.3]$, avoiding phase-separated states that appear for $\phi\geqslant0.4$ \cite{fily2012athermal}. We performed our measurements over 30 independent realizations of the initial conditions. In each realization, the total
number of time steps is $2\times 10^{6}$, measurements are made in the last $10^6$ steps, and an integration time step $h=10^{-3}$.







\section{Results and Discussions}
\label{resultAnd}


We divide this section in three distinct subsections is order to present more clearly our results and interpretations: in the first one, we show general features of the correlated states and the typical motion pattern of the vortices in each state; next, we show the dependence of the correlation function on $\delta$ and $D$; in the last one, we discuss the uncorrelated state, which occur at the boundary region between the positive and negative correlated regimes and for large gaps.
\begin{figure}[ht]	
 \includegraphics[width=\linewidth]{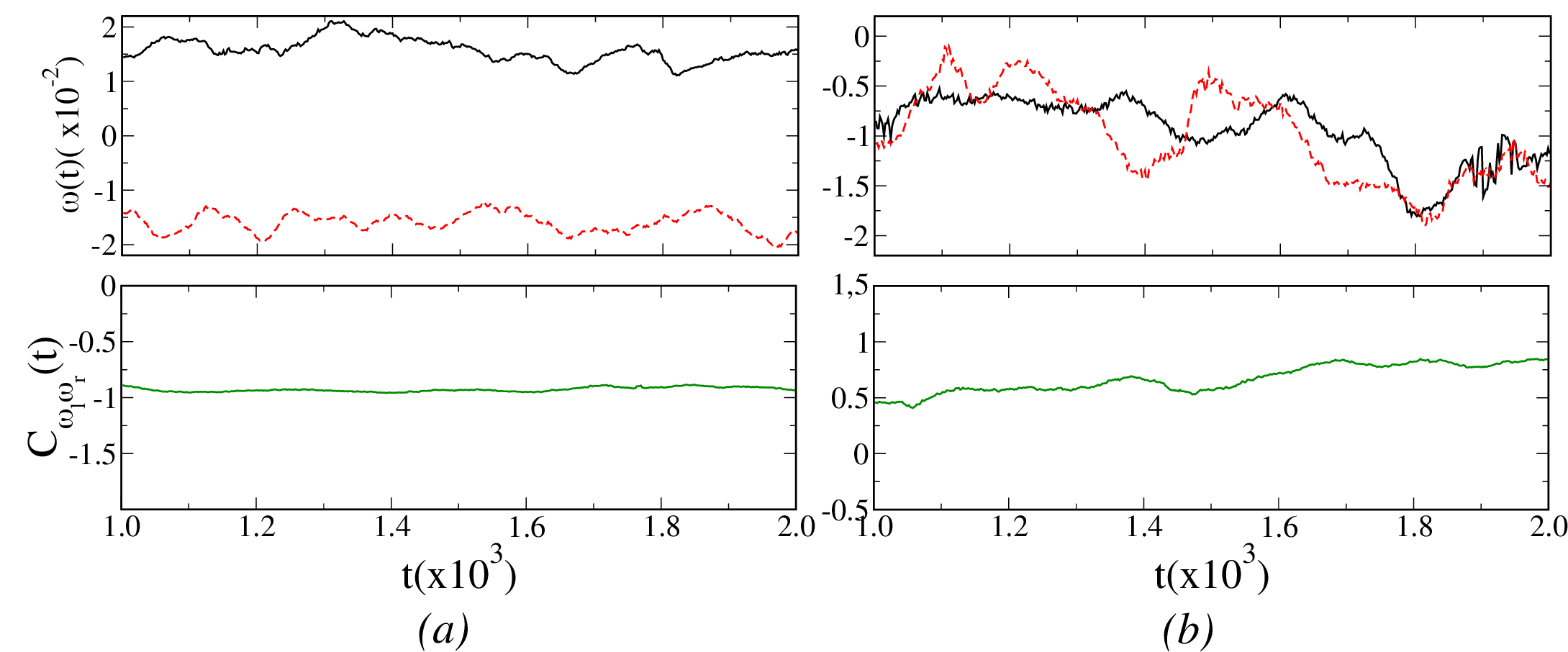}

\caption{Angular velocity as a function of time, for a single run, of the left vortex ($\omega_{l}$), black curves, and the right vortex ($\omega_{r}$), red curves (upper panels), and the corresponding correlation function averaged over independent runs (lower panels) for (a) a NCS at $\delta=10$, $D=40$, $\phi=0.25$, and (b) a PCS at $\delta=2$, $D=25$, $\phi=0.30$.}
\label{fig:OmegaT}
\end{figure}
\subsection{General features of the correlated states\label{subsec3-1}}
\begin{figure}[ht]	
    \centering
\includegraphics[width=0.85\linewidth]{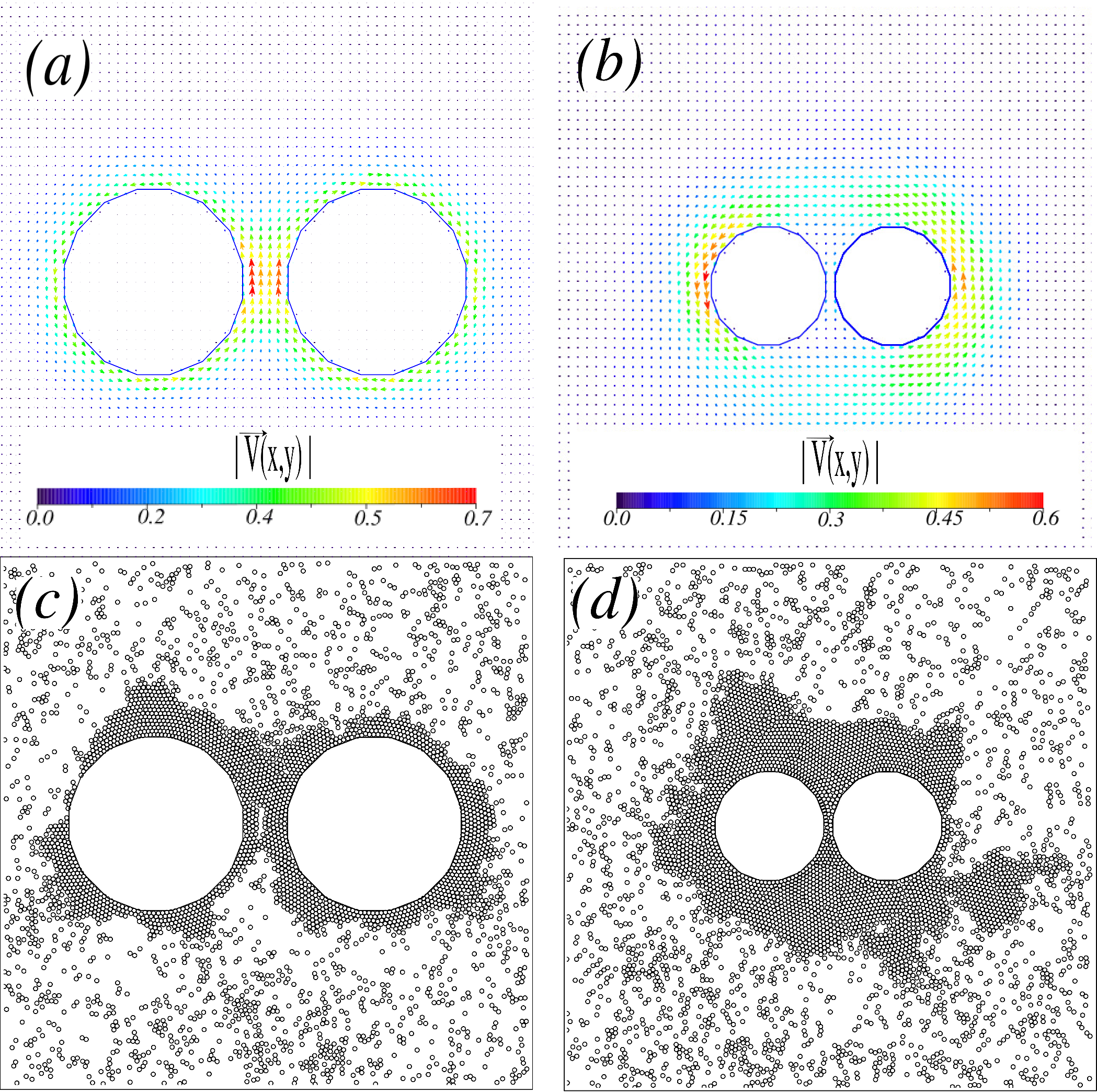}
  
 	\caption{(a) The velocity field averaged over time for a single run for $\phi=0.25$, $\delta=10$ and $D=40$; (b) $\phi=0.30$, $\delta=2$ and $D=25$. Typical particle configurations for (c) a NCS and (d) a PCS.}
 	\label{fig:Top12}
 \end{figure}

We start by presenting the general features of the correlation function. 
In Figs. \ref{fig:OmegaT}(a) and (b), we show, in the upper panels, the angular velocities of each vortex as a function of time for a single run, and in the lower panels, we show the corresponding correlation function as a function of time, averaged over $30$ independent realizations. In Fig. \ref{fig:OmegaT}(a), $C_{\omega_{l}\omega_{r}}(t)<0$ indicates that the vortices rotate in opposite directions. We also observe that the magnitude of angular velocities are equivalent (such equivalence is seen in all runs). This states is a NCS. In Fig. \ref{fig:Top12}(a), we show a typical velocity field and in Fig. \ref{fig:Top12}(c), a typical particle configuration for the NCS. Conversely, in Fig. \ref{fig:OmegaT}(b), $C_{\omega_l\omega_r}(t)>0$ indicates that the vortices rotate in the same direction, {\em i.e.}, a PCS, with $|\omega_{l}|\approx|\omega_{r}|$ as in the previous case. In Fig. \ref{fig:Top12}(b), we show a typical velocity field and in Fig. \ref{fig:Top12}(d), we show a particle configuration for this state. 

There is a third possible outcome for the correlation function, which we omit here, namely the uncorrelated state, $C_{\omega_{l}\omega_{r}}\approx 0$. 
This case is usually observed when the vortices are widely separated. In this regime, as we will show, in some cases, the correlation function does not vanish exactly, but the motion patterns of the correlated state are distinct to those seen in Figs. \ref{fig:Top12}(a) and (b), which lead us to define a condition that unambiguously characterizes whether a state is correlated or not. Finally, we also observe uncorrelated states in the transition region between the PCS and the NCS. In this case, usually only one vortex is present, which renders a nearly vanishing  correlation.

The velocity fields and configurations shown in Fig. \ref{fig:Top12} give us an indication of the shapes and motion patterns of the vortices in both correlated states. In the NCS, Figs. \ref{fig:Top12}(a) and (c), both vortices are separated, visually distinguishable, and exhibit the typical isolated rotation pattern around each corresponding obstacle. However, they overlap in the inner region between them, where the vertical particle current is stronger than the vortices mean tangential velocities. The emergence of the NCS can be understood under the capture-and-release mechanism proposed in \cite{mokhtari2017collective,pan2020vortex} as follows. As particles adhere to an obstacle's surface, they form a rough layer of particles that captures additional particles, increasing the aggregate. By natural fluctuations, as particles rotate about the obstacle, one of the rotation directions will prevail over the other and a vortex will appear about the obstacle. In the present case, this will happen for both obstacles. However, they will hardly arise simultaneously. When one of them is formed, it will provide a continuous flow of particles to the other obstacle. These particles, for instance, will move downwards (upwards) for a clockwise (counterclockwise) left vortex. Now, this flow will supply the incoming particles that will eventually overcome the influence of the natural fluctuations on the other vortex, and will drive the other vortex in a direction contrary to the first one. Hence, a NCS ensues from this particle exchange. This motion pattern resembles two synchronized cogs, and we call it cog-like correlation. That is, the negative correlation can be linked to two synchronized cogs rotating in contrary directions, exchanging particles between them. 

On the other hand, in the PCS, Figs. \ref{fig:Top12}(b) and (d), the vortices merge and rotate around both obstacles as a single structure, with the inner current lower than the mean tangential velocity. 
(See the supplemental material for videos of both states). 
Qualitatively, in the density range in which we worked, there are enough particles to accumulate between the obstacles, and which are essentially immobile, while a significant amount of them remains free to move. The other particles, therefore, are able to slide over the gap and circumvent both obstacles. Again, within the capture-and-release picture of \cite{mokhtari2017collective,pan2020vortex}, a single vortex emerges because there is, effectively, a single obstacle. This motion pattern is similar to a belt strapped around both obstacles with particles orbiting the obstacles as a single unit. We call that a belt-like correlation. This scenario should be contrasted to the ones seen in \cite{harder14,ray14,leite16,Casimir_Forces_Rohwer,C8SM01840E,D0SM01797C}, in studies of Casimir forces between two static obstacles, in which particles also accumulate between the obstacles, but no belt-like motion is seen due to the very low density that was used, typically $\phi\leq0.1$. This is an evidence that correlated vortices emerge above a certain density. 

\subsection{Correlation dependence on $D$, $\delta$, and $\phi$\label{subsec3-2}}
It was shown \cite{pan2020vortex,mokhtari2017collective} that the rotation of a single vortex mainly depends on the obstacle diameter. We state that, for two obstacles, the gap, as well as the particle density, can also be important factors that determine correlations. Therefore, we present the results of the dependence of the correlation function on the obstacle diameter $D$, gap $\delta$, and particle density $\phi$.

\begin{figure}[ht]	
	\includegraphics[width=\linewidth]{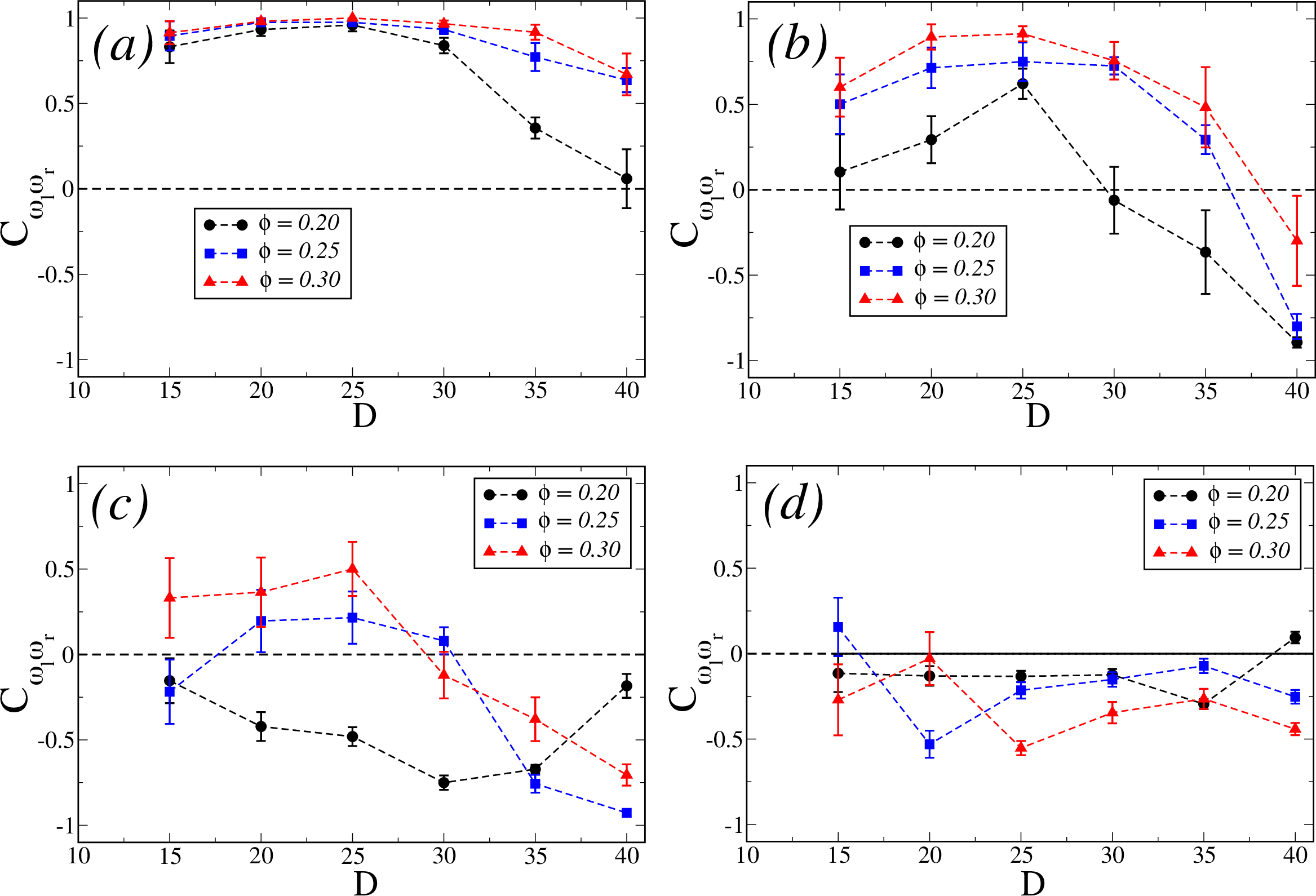}
	\caption{The correlation function  ($C_{\omega_{l},\omega_{r}}$) as a function of obstacle diameter ($D$), for different  $\phi$ for: (a) $\delta=2$; (b) $\delta=4$; (c) $\delta=10$ and (d) $\delta=20$.}
	\label{fig:Top11}
\end{figure}

In Fig. \ref{fig:Top11}, we show the correlation function  $C_{\omega_{l}\omega_{r}}$ as a function of the diameter of the obstacles $D$ for different $\delta$ and $\phi$. For the smallest gap, $\delta=2$, Fig. \ref{fig:Top11}(a), we observe the PCS for all $\phi$, with a decreasing $C_{\omega_{l}\omega_{r}}$ as $D$ increases, eventually reaching nearly zero correlation for $\phi=0.20$ and $D=40$. We also see that for $D\leqslant 25$, the correlation depends weakly on $\phi$, while, for $D>25$, only the $\phi=0.2$ case shows a strong decrease with $D$. 

At $\delta=4$ Fig. \ref{fig:Top11}(b), we observe the reduction of the intensity of the correlation in the PCS and the emergence of the NCS, as compared to the $\delta=2$ case, mainly for $\phi=0.20$ and $D\geqslant 30$. For larger $\phi$, it appears only for $D=40$. Finally, unlike the previous case behavior with $\phi$, increasing density results in an overall increase of correlation at fixed $D$.
  
In Fig. \ref{fig:Top11}(c), we show the results for $\delta=10$. We observe both correlated states, but now the NCS is predominant, in agreement with the trend observed when we passed from the  $\delta=2$ to the $\delta=4$ cases. For $\phi=0.20$, the correlation is negative for any $D$, while for $\phi>0.20$, this regime occurs only for $D\geqslant 30$. For smaller obstacles, we still observe a positive, albeit small, correlation.

Finally, in Fig. \ref{fig:Top11}(d), we present the results for $\delta=20$. Here, the NCS occurs in almost all cases. Also, we see that the correlation function has a weaker dependence on $\phi$ compared to the $\delta=10$ results.

\begin{figure}[ht]
 	\centering
 	\includegraphics[width=\linewidth]{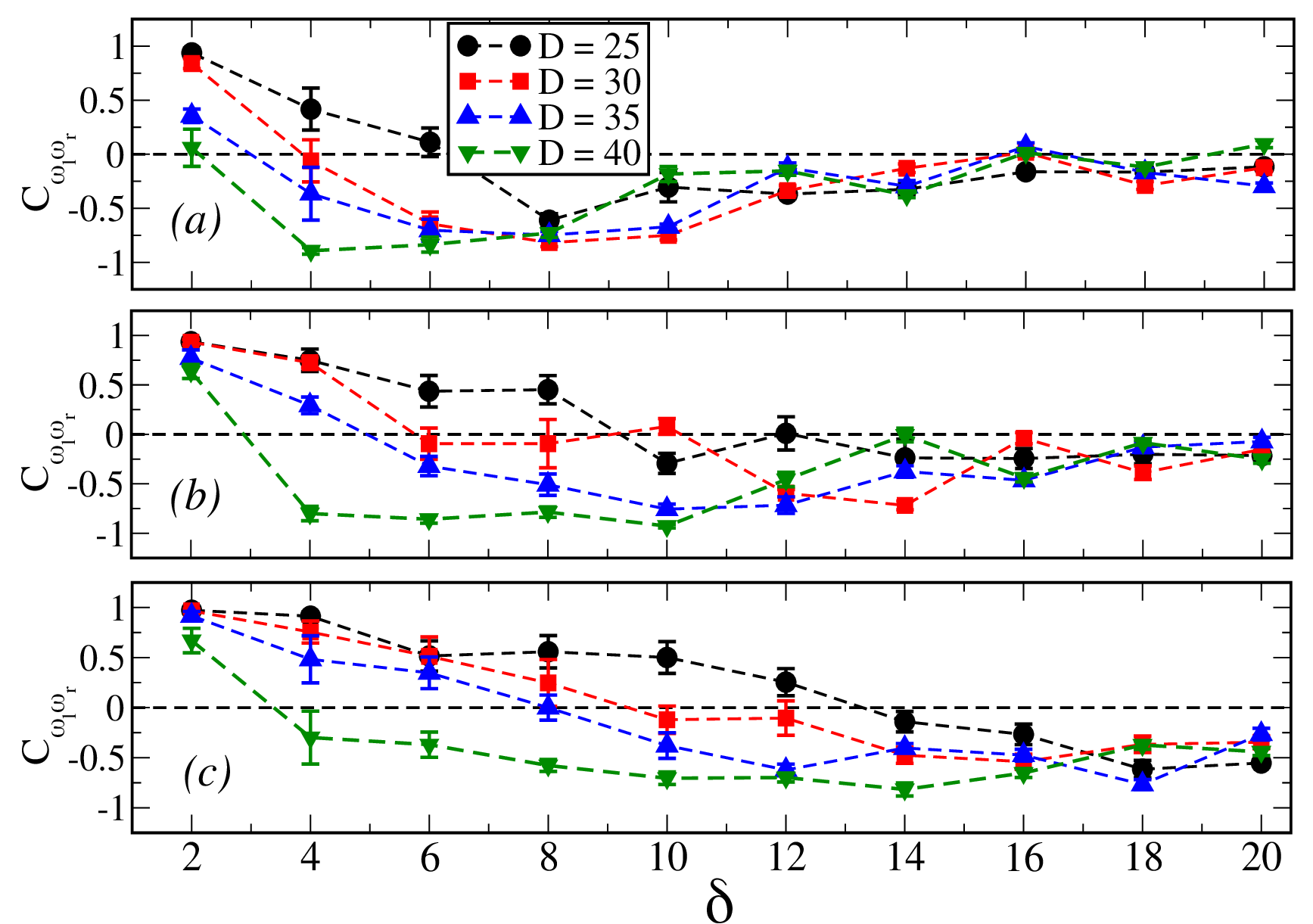}
 	\caption{Correlation function  $C_{\omega_{l}\omega_{r}}$ as a function of gap $\delta$ for distinct obstacle's diameter $D$, for particle density (a) $\phi=0.20$; (b) $\phi=0.25$; and (c) $\phi=0.30$.}
 	\label{fig:correlationbydelta}
 \end{figure}

 From these data, we observe a general decrease of the correlation with increasing $D$, this trend being more pronounced for $\phi=0.30$. For the other two density values, the $\delta$ range in which we see this behavior is shorter for lower $\phi$. For instance, for $\phi=0.20$ the correlation function decreases with $D$ clearly only up to $\delta=4$, while for $\phi=0.25$, only up to $\delta=10$. We also see that, at $\delta=20$, the correlation function is nearly insensitive to increasing $D$. The belt and cog analogies introduced in sec. \ref{subsec3-1} are helpful in interpreting these results. In the PCS, as we increase the obstacles' diameter, but keep the gap and density fixed, we increase the space between the obstacles, and more particles are needed to fill this region which would allow the other to move about both obstacles (in studies of trapping of active matter \cite{kaiser2012} such state was called a partial trapping state); also the particles should run a longer path in order to circumvent both obstacles, effectively stretching the belt, {\em i.e.}, producing a vortex with less particles than the one at lower $D$. This effect decreases the correlation function, since it would take more particles to form a belt at the same effective density as that at lower $D$. In the NCS, a somewhat opposite effect takes place: as we increase $D$, while keeping $\delta$ and $\phi$ fixed, more particles are captured by the vortices, since the surface that they stick to are larger. These captured particles will eventually reach the region in which the vortices exchange particles, resulting in larger cogs that would reach out to each other for the correlation to occur. Of course, this is a limited effect because increasing $D$, at fixed $\delta$ and $\phi$, indefinitely will produce a very dilute system in which a single vortex cannot be sustained. Next, we pass on to the discussion of the influence of the gap on the correlation function and show the dependence of $C_{\omega_l\omega_r}$ on $\delta$, Fig. \ref{fig:correlationbydelta}.
 
As seen in Fig. \ref{fig:correlationbydelta}(a), for $\phi=0.20$, the decrease in correlation intensity with increasing $\delta$ is more clearly observed, especially for $\delta\leqslant 8$, for any $D$. For larger gaps, the correlation approaches zero and shows a weak dependence on both $D$ and $\delta$. For $\phi=0.25$, Fig. \ref{fig:correlationbydelta}(b), the initial decreasing trend of $C_{\omega_l\omega_r}$ with $\delta$ is seen up to $\delta=14$, while above this gap value, the correlation depends weakly on $D$ and $\delta$. Finally, at $\phi=0.30$, Fig. \ref{fig:correlationbydelta}(c), this decreasing trend lasts up to $\delta=18$. In general, these curves reach a lowest correlation point at a gap value that depends on $D$ and $\phi$; for larger gaps, the states have small negative or null correlations for $\phi=20$ and $0.25$; for $\phi=0.30$, this last regime does show relevant correlations.

The observed trend in these results is that increasing the gap, while keeping $D$ and $\phi$ constant, generally changes the correlated state from a PCS to a NCS. We can understand this scenario through the belt and cog analogies as above. Let us consider initially the PCS, such as the one depicted in Figs. \ref{fig:Top12}(b) and (d). We see that as we increase the gap, at fixed $D$ and $\phi$, the space between the obstacles increases and the particles should travel a longer path enclosing both obstacles. Hence, by the same reasoning used above, the belt necessarily stretches. Then, if we continue this process, it will reach a snapping point, in which the particles will no longer be able to move around both obstacles, {\em i.e.}, we will reach the transition region. The approach to this region is continuous, since the correlation function decreases continuously with increasing gap. By increasing this parameter even further, nearly separated vortices will appear around each obstacle, as seen in Figs. \ref{fig:Top12}(a) and (c). At this point, the NCS arises. 

A question remains as to whether the vortices, whose correlations are negative for $\delta$ above the lowest correlation point, see Fig. \ref{fig:correlationbydelta}, are indeed correlated, or merely slightly perturbed by the presence of the other vortex. This is markedly important for larger obstacles $D\geqslant 35$, since we expect them to be in the stable vortex state, in which they rarely change their rotation directions throughout the simulation. From that, we may not be able to distinguish between two counter-rotating vortices that occurred by chance from a true NCS. We will address this issue below in sec. \ref{subsec3-3}. Before that, we comment on the dependence of the correlation on $\phi$.

Based on the observation that active particles tend to stick to solid surfaces, increasing $\phi$ results in more particles captured by vortices \cite{mokhtari2017collective,pan2020vortex}, making them more dense. In terms of the belt and cog analogies, we may say that for more dense systems, the belts and cogs will be \emph{thicker}, meaning that a larger gap is required to reach the belt snapping point and the lowest correlation point as compared to those at lower $\phi$. Also, we observed,  for $\phi<0.20$, that the correlations are lost in such diluted regimes since the vortices have too few particles to produce any correlated state (again, compare to the states seen in \cite{harder14,ray14,leite16,Casimir_Forces_Rohwer,C8SM01840E,D0SM01797C}). On the other hand, increasing the density even further than $0.30$ will decrease the correlations, and a correlated state will not be observed, since a dense structure is formed around both obstacles that prevents any rotation. In other words, in both cases, there are no observable rotating vortices.

\subsection{Uncorrelated states: transition region and large gap regime\label{subsec3-3}}
Now, we comment on the large gap regime, {\em i.e.}, states at $\delta$ values that are above the one in which the lowest correlation is observed. We see from Figs. \ref{fig:Top11} and \ref{fig:correlationbydelta} that there is the predominance of negatively correlated states, but the correlation is, mostly, small. This feature is more evident for $\phi\leq0.25$ at all $D$. There are also a few points in Figs. \ref{fig:correlationbydelta}(a) and (b) that have nearly vanishing correlation. For $\phi=0.30$, however, the correlations at $\delta\geq18$ are still comparable to those at lower gaps.  We claim that some of these states are effectively uncorrelated because, even though we do see a finite negative correlation, the motion pattern of the particles is not the typical one for a NCS, see Fig. \ref{fig:Top12}(a). More specifically, the vertical particle current between the obstacles is lower than the tangential velocities of both vortices. In order to make this distinction more clearly,
we can define an uncorrelated state by the general shapes of the vertical velocity horizontal profiles, $J_y(x)$ (in fact, this provides an easier way to observe this distinction compared to direct visualization of the velocity fields, but both data sets yield the same results). We see that each state has a particular signature for this profile.

\begin{figure}[ht]	
 \centering
 \includegraphics[width=\linewidth]{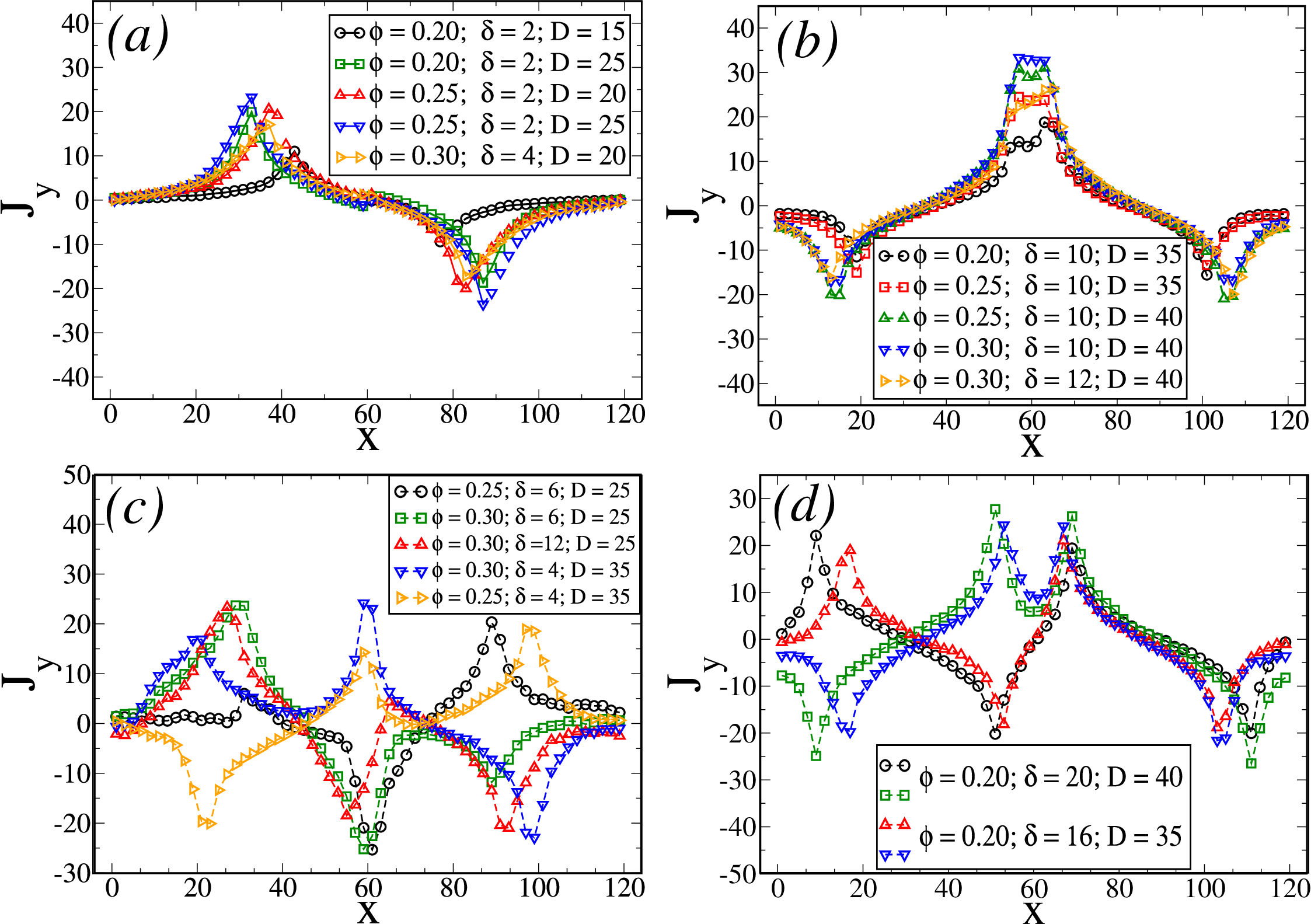}
	\caption{The mean vertical particle current profiles as a function of the horizontal position $x$ for (a) PCS and (b) NCS. In (c), we show five single run profiles for states in the transition region between the PCS and the NCS. In (d), we show four single run profiles for uncorrelated states in the large gap regime.}
	\label{fig:profile}
\end{figure} 

In Fig. \ref{fig:profile}, we show the mean vertical velocity horizontal profiles measured directly from the velocity fields for all four possible correlated states: PCS, Fig. \ref{fig:profile}(a), NCS, Fig. \ref{fig:profile}(b), uncorrelated, Fig. \ref{fig:profile}(c), and low correlation, Fig. \ref{fig:profile}(d). We see that the PCS, Fig. \ref{fig:profile}(a), exhibit two peaks of nearly the same magnitude but of opposite signs, and occurring nearly at the outermost vertical tangent of the obstacles, {\em i.e.}, at $x_{l,r}=L/2\mp(D+\delta/2)$, the minus(plus) sign refers to the left(right) obstacle. Specifically, the contrary peaks occur at the leftmost point of the left obstacle and at the rightmost point of the right obstacle, and indicates a single rotating structure, or a belt around both obstacles, since at the center of the system, $x=L/2$, the vertical current nearly vanishes. For the NCS, Fig. \ref{fig:profile}(b), there are three peaks: two occur, again, at $x_{l,r}$, and are nearly of the same magnitude and have the same sign, and a third peak situated at the center of the system. This third peak is broader than the others, spanning the entire gap, with a distinct sign compared to the other peaks and larger in magnitude. This indicates the presence of a strong particle current in the space between the vortices, stronger than the tangential velocities of the isolated vortices. As aforementioned, it is formed by particles in both vortices that travel in the same direction in that region. 

For the uncorrelated states, Fig. \ref{fig:profile}(c), there are usually three peaks in each of them; two of them, one at the center $x=L/2$ and the other at either $x_l$ or $x_r$, have nearly the same magnitude but contrary signs. For instance, see the blue triangles curve: it indicates that the there is a clockwise vortex around the right obstacle; its third peak is at $x_l$ and it has the same sign as the center one. Therefore, at the left obstacle, particle in the center move in the positive $y$ direction, implying a counterclockwise vortex, while those at $x_l$ also move in the positive $y$ direction, indicating a clockwise vortex. Since this configuration amounts to distinct rotation directions at opposite side of the left obstacle, we conclude that there is no vortex around this obstacle. These features can be seen in all other profiles. These states are in the transition region between the PCS and the NCS and correspond to the belt-snapping point we mentioned earlier.

Finally, for the low correlation states, Fig. \ref{fig:profile}(d), let us focus on the black circles and red triangles. There are four peaks in each of them, two for each obstacle. Both profiles indicate a positive correlation. Nevertheless, compare these profiles to the ones shown in Fig. \ref{fig:profile}(a) for the single-vortex PCS. It is clear now that we have two separated vortices rotating in the clockwise direction instead of a single one enclosing both obstacles. Now, let us focus on the green squares and blue inverted triangles. They have similar shapes, but are reflected at the center of the system resulting in a negative correlation. Again, compare these profiles to those seen in Fig. \ref{fig:profile}(b) for the two-vortex NCS: it is evident that both vortices exchange particles through the center of the system, but the particle current is lower than the peak of the tangential velocities at $x_{l,r}$. We can consider this state to be the one in which the cogs no longer touch each other and can rotate freely. 

\begin{figure}[ht]	
	\centering
 \includegraphics[width=\linewidth]{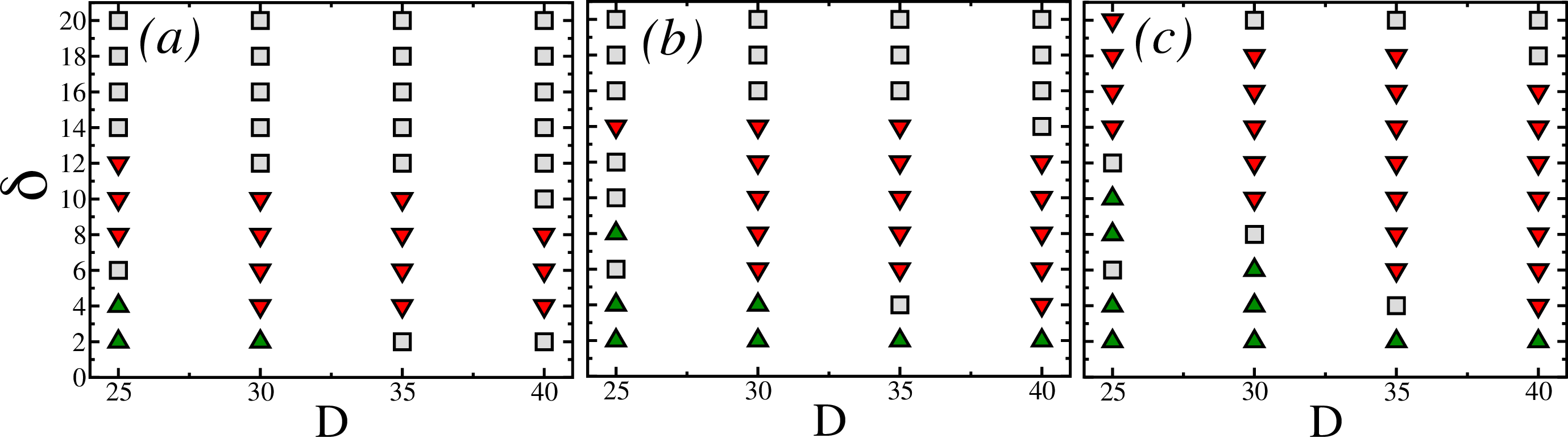}
	\caption{ Phase diagrams $\delta$ vs.$D$ for $D\geqslant25$ at (a) $\phi=0.20$; (b) $\phi=0.25$; and (c) $\phi=0.30$. Red inverted triangles represent the NCS, grey squares, uncorrelated states, and green triangles, the PCS.}
	\label{fig:Top1Phase}
\end{figure}

By analyzing all profiles, we generated in Fig. \ref{fig:Top1Phase} a set of $\delta$ vs. $D$ phase diagrams for different $\phi$ indicating the observed correlations. We defined two parameters for the determination of a correlated state. To define a NCS, we first check whether the correlation function is negative; then, from the vertical particle current profiles for each run, we calculate the following ratio for the $i$-th run ($i=[1,...,30]$, the total number of independent runs)
\begin{equation}
    \label{eqn10} 
    \mathcal{A}_i=\frac{2|J_y(L/2)|_i}{|J_y(x_l)|_i+|J_y(x_r)|_i},
\end{equation}
which is the one between the vertical velocity at the center of the system to the mean vertical velocity at $x_{l,r}$. If the mean value of this ratio over all runs, $\mathcal{A}=1/30\sum_i\mathcal A_i$, is larger than $0.5$, we consider this state to be a NCS. To define a PCS, we observe whether the correlation function is positive; and from the vertical velocity profiles, we calculate the ratio for the $i$-th run
\begin{equation}
    \label{eqn11}
\mathcal{F}_i=\frac{|J_y(x_l)+J_y(x_l+D)|_i+|J_y(x_r)+J_y(x_r-D)|_i}{|J_y(x_l)|_i+|J_y(x_r)|_i},
\end{equation}
which is the one between the sum of the vertical velocities of both vortices at their outermost and the innermost vertical tangents to the sum of the vertical velocities at $x_{l,r}$. In this case, if the mean value of $\mathcal{F}_i$ over all runs, $\mathcal{F}=1/30\sum_i\mathcal{F}_i$, is also larger than $0.5$, the system is considered to be a PCS. These criteria are chosen so that the resulting phase diagrams reflect the data show in Figs. \ref{fig:Top11} and  \ref{fig:correlationbydelta} for the correlation function. In all other cases, we consider the system to be uncorrelated. Notice, in Fig. \ref{fig:Top1Phase}, that the NCS is more common than the PCS, and that the latter occurs more frequently at $D=25$, suggesting that the belt is more easily formed for smaller obstacles. Additionally, increasing density enlarges the parameter range in which both states appear, while decreases the parameter range of the uncorrelated states in agreement with the analogy of thicker belts and cogs stated earlier, see sec. \ref{subsec3-2}.

\section{CONCLUSIONS}
\label{CONCLUSIONS}

We investigated the dynamics of dry active matter in the presence of two identical circular obstacles and focused in the correlations between the rotations of the vortices formed around each obstacle. We measured the angular velocity correlation function between the vortices as a function of the obstacle size $D$, gap $\delta$, and particle density $\phi$. We observed two distinct regime regarding the vortices' motion around the obstacles: a positive correlation regime characterized by the merging of the two vortices in a single rotating cluster around both obstacles that resembles a belt strapped around them. This regime is primarily observed for low gap and obstacle size.
A negative correlation regime composed of two distinct counter-rotating vortices which exchange a significant amount of particles in the region between them, resulting in a non-vanishing, strong vertical particle current (compared to the vortices' tangential velocities) at the center of system. This particles current keeps both vortices' motions correlated. This behavior resembles two synchronized cogs, and is typically observed for intermediate gap and large obstacle size. 

We also observed the occurrence of uncorrelated states in which both vortices are approximately independent of one another, even though the correlation function does not vanish in a typical run. These uncorrelated states occur mainly for large gap values and low obstacle sizes, we consider them to be the point in which the cogs formed by the vortices no longer touch each other and rotate freely. 

We also observed uncorrelated states in which only one vortex is present. This corresponds to the belt snapping point and marks the transition between the two correlation regimes.

We defined two conditions, eqs. (\ref{eqn10}) and (\ref{eqn11}), based on the comparison of the magnitudes of the tangential vertical velocities of the vortices, in order to determine whether a given state is a NCS, a PCS or an uncorrelated state. From these conditions, we were able to build a phase diagram that summarizes and reflects the data we present throughout the text.


Finally, increasing the particle density strengthens the correlation. In other words, when we increase the density, the ranges of $\delta$ and $D$ in which both correlated states occur become larger. We interpreted this result as a strengthening of the vortices: they capture more particles allowing the obstacles to be placed farther apart or to be larger, while still exhibiting one of the two correlated motion states.

A possible future direction for this investigation is the extension of the current setup to a regular lattice of circular obstacles in order to explore whether any synchronized state, similar to those observed for wet models \cite{wioland2016ferromagnetic,reinken2022ising}, emerges. This is actually a challenging problem because achieving an uniform distribution of particles throughout the lattice is rather difficult, given the clustering tendency observed in dry active matter. Alternatively, we may observe pairs of correlated vortices in a lattice of obstacles, but their behavior would likely resemble what we have just described.





\bibliographystyle{elsarticle-num}
\end{document}